\RequirePackage{fix-cm}
\documentclass[smallextended]{svjour3}       
\smartqed  
\usepackage{appendix}
\usepackage{amsmath}
\usepackage{graphicx}
\usepackage{lineno}
\usepackage{array}
\usepackage{longtable}
\usepackage{natbib}
\newcommand*\patchAmsMathEnvironmentForLineno[1]{%
\expandafter\let\csname old#1\expandafter\endcsname\csname #1\endcsname
\expandafter\let\csname oldend#1\expandafter\endcsname\csname end#1\endcsname
\renewenvironment{#1}%
{\linenomath\csname old#1\endcsname}%
{\csname oldend#1\endcsname\endlinenomath}}%
\newcommand*\patchBothAmsMathEnvironmentsForLineno[1]{%
\patchAmsMathEnvironmentForLineno{#1}%
\patchAmsMathEnvironmentForLineno{#1*}}%
\AtBeginDocument{%
\patchBothAmsMathEnvironmentsForLineno{equation}%
\patchBothAmsMathEnvironmentsForLineno{align}%
\patchBothAmsMathEnvironmentsForLineno{flalign}%
\patchBothAmsMathEnvironmentsForLineno{alignat}%
\patchBothAmsMathEnvironmentsForLineno{gather}%
\patchBothAmsMathEnvironmentsForLineno{multline}%
}

\usepackage{color}

\begin{document}

\title{Effect of the land surface thermal patchiness on the Atmospheric Boundary Layer through a quantification of the dispersive fluxes.}

\titlerunning{Surface thermal patchiness, the ABL, and dispersive fluxes}        

\author{Fabien Margairaz    \and
        Eric Pardyjak       \and
        Marc Calaf
}
\authorrunning{Margairaz et. al} 

\institute{
    Fabien Margairaz \at
        Department of Mechanical Engineering, University of Utah, Salt Lake City, Utah, USA\\
        \email{fabien.margairaz@utah.edu} 
    \and
    Eric Pardyjak \at
        Department of Mechanical Engineering, University of Utah, Salt Lake City, Utah, USA\\
    \and
    Marc Calaf \at
        Department of Mechanical Engineering, University of Utah, Salt Lake City, Utah, USA\\
        \email{marc.calaf@utah.edu}           
}

\date{Received: DD Month YEAR / Accepted: DD Month YEAR}
\maketitle

\begin{abstract}
While advances in computation are enabling finer grid resolutions in numerical weather prediction models, representing land-atmosphere exchange processes as a lower boundary condition remains a challenge. This partially results of the fact that land-surface heterogeneity exists at all spatial scales and its variability does not `average' out with decreasing scales. The work here presented uses large-eddy simulations and the concept of dispersive fluxes to quantify the effect of surface thermal heterogeneity with scales $\sim$ 1/10th the height of the atmospheric boundary layer and characterized by uniform roughness. Such near-canonical cases describe inhomogeneous scalar transport in an otherwise planar homogeneous flow when thermal stratification is weak or absent. Results illustrate a regime in which the flow is mostly driven by the surface thermal heterogeneities, in which the contribution of the dispersive fluxes can account for more than 40\% of the total sensible heat flux at about 0.1$z_i$, with a value of 5 to 10\% near the surface, regardless of the spatial distribution of the thermal heterogeneities, and with a weak dependence on time averaging. Results also illustrate an alternative regime in which the effect of the surface thermal heterogeneities is quickly blended, and the dispersive fluxes match those obtained over an equivalent thermally homogeneous surface. This character seems to be governed by a new non-dimensional parameter representing the ratio of the large scale advection effects to the convective turbulence. We believe that results from this research are a first step in developing new parameterizations appropriate for non-canonical atmospheric surface layer conditions.
\end{abstract}

\section{Introduction}\label{sec:intro}
 
While advances in computation (and computing) are enabling finer grid resolutions in numerical weather prediction (NWP) models, representing land-atmosphere exchange processes as a lower boundary condition remains a challenge (regardless of the numerical resolution but not independent from it). This is partially a result of the fact that land-surface heterogeneity exists at all spatial scales and its variability does not `average' out with decreasing scales. Such variability need not rapidly blend away from the boundary thereby impacting the atmospheric surface layer (ASL). The correspondingly induced mixing and forcing processes are characterized by short time scales ($\sim \mathcal{O}(1 \rm{hour})$) and limited spatial extent ($\sim \mathcal{O}(100\rm{m})$) that fall within the so-called `terra-incognita' or `gray-zone' range \citep{Wyngaard2004,Beare2014}, making it difficult for traditional NWP models to capture them. 

\begin{figure}
  \begin{center}
    \includegraphics[width=0.7\textwidth]{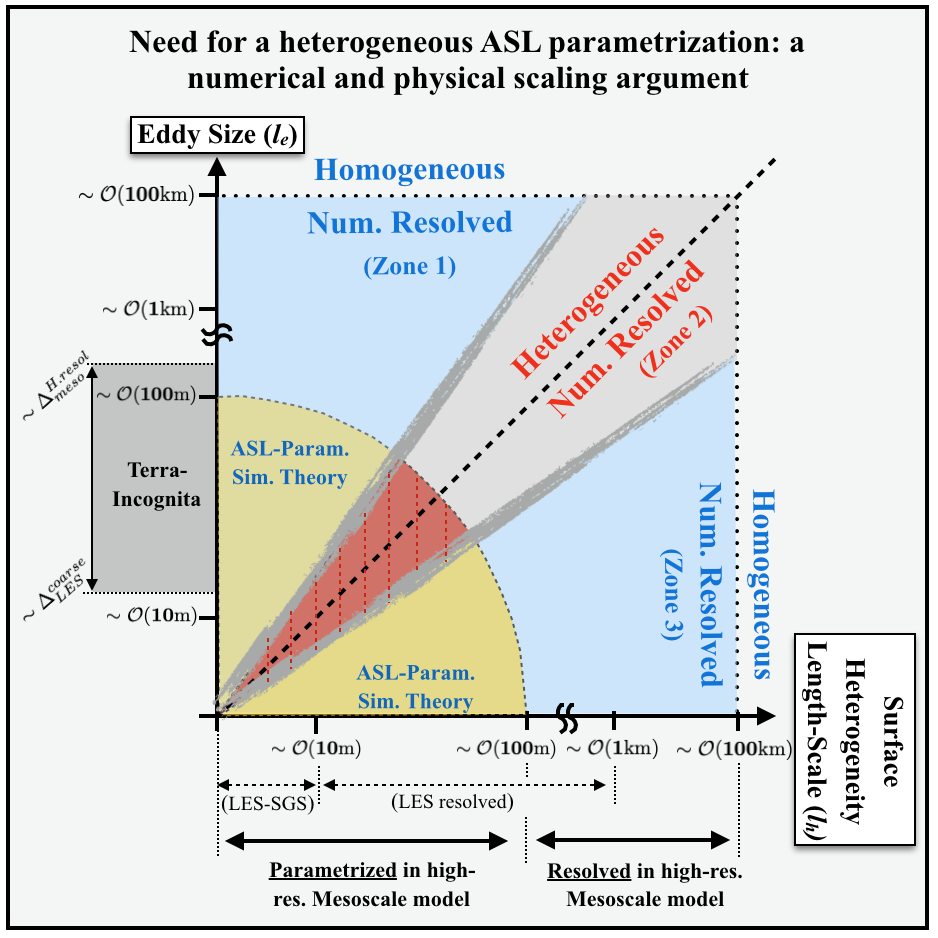}
  \end{center}
  \caption{\footnotesize{Illustration of the interaction between turbulent eddies and heterogeneous surface patches of different scales to motivate the need for new ASL parameterizations in mesoscale modeling.}}
  \label{fig:LengthScales}
\end{figure}

Figure~\ref{fig:LengthScales} presents a conceptual scheme that illustrates the relationship between the different flow turbulent scales, the land surface heterogeneity scales and the available numerical resolution, which has been used to guide the development of this work. In Figure~\ref{fig:LengthScales}, the abscissa represents horizontal surface heterogeneity length scales ($l_h$), and the ordinate the size of turbulent eddies ($l_e$). Four relevant length-scales are represented on both coordinate axes, $\sim \mathcal{O}(100 \rm{m})$ represents the smallest turbulent scales resolved in near-future high-resolution mesoscale models ($\Delta_{meso}^{H.resol}$), and $\sim \mathcal{O}(100 \rm{km})$ the horizontal domain size ($L_{meso}$). Additionally, $\sim \mathcal{O}(10 \rm{m})$ and $\sim \mathcal{O}(1 \rm{km})$, represent the same corresponding length scales for high-resolution LES ($\Delta_{LES}^{H. resol}$ and $L_{LES}$). Moreover, turbulent scales are identified as either numerically resolved or parameterized (different color-shading). An arc is placed at scales $\sim \mathcal{O}(100 \rm{m})$. Within this arc and above the diagonal are regions where unresolved (in mesoscale models) eddies are much larger than the scales of heterogeneity ($l_e/l_h>>1$). Here, turbulence feels the surface as homogeneous and traditional ASL formulations should be valid. Within the arc and below the diagonal, unresolved eddies are much smaller than the scales of heterogeneity ($l_e/l_h<<1$) and again the turbulence feels the surface as homogeneous and traditional ASL formulations should be valid. Within an uncertain zone along the diagonal, eddy scales are of the same order as scales of heterogeneity (i.e., $l_e \sim l_h$). It is precisely this heterogeneous numerically unresolved zone that frames the scope of this work. Much of this zone overlaps with anticipated increases in NWP resolution as well as the so-called `terra-incognita' or `gray-zone' \citep{Wyngaard2004,Beare2014}. It is also the zone that is most problematic for ASL similarity relations \citep{Patton2005,Li2013} and has been known as a source of error that is introduced by the indirect \textit{homogenization} of surface spatial heterogeneities. This error depends on the resolution of the numerical model, the characteristic length-scale of the surface heterogeneities, and the parameterization used for the near-surface conditions \citep{Patton2005,Li2013}.

Because the flux-similarity relationships were developed under the assumption of spatial homogeneity and statistical stationarity \citep{MoninObukhov54,Stull88}, prior efforts have attempted to overcome this inherent limitation using more or less sophisticated approaches to provide adjustments to the parameterization of the Reynolds flux component of the sensible heat flux. Some example of these strategies are the effective surface parameter approach \citep{Wieringa1986, Taylor1987, Mason1988, Wood1991, Beljaars1991, Claussen1990, BouZeid2004}, the statistical-dynamical approach \citep{Avissar1991,Avissar1992}, and the mosaic and the tile approaches \citep{Avissar1989, Claussen1991, Ament2006} or any modification of the ones above such as the extended tile \citep{Blyth1993,Blyth1995, Arola1999}. These methods are currently used in most NWP and climate models to parameterize surface energy fluxes over heterogeneous terrain because they are easy to implement and use, and due to the current lack of better alternatives. Further, because both the Tile and the Mosaic approaches were developed to be applied above the blending height, the framework under which these should be applied will become more severely challenged as numerical resolution increases. Additional compelling evidence of the limitations of these approaches results from the consistent overestimation produced by flux-profile methods \citep{Stoll2009} in most heterogeneous cases. 

As a first step towards developing new ASL parametrizations for high-resolution NWP models that overcome the problems presented above, the contribution of the dispersive fluxes is evaluated on flows over idealized thermally-heterogeneous surfaces with uniform roughness. Note that the dispersive fluxes, as identified in earlier works \citep{Wilson1977, Raupach1982, Finnigan1985, Raupach1986}, are terms arising from the spatial averaging operation that explicitly represent the critical processes dependent on heterogeneity. Therefore, it is here hypothesized that when thermal spatial homogeneity is not satisfied, the contribution of the dispersive fluxes becomes important. In Section \ref{sec:theory} the derivation and theory related to the dispersive fluxes is reviewed; in Section \ref{sec:Simul_cases} the numerical platform used and study cases are presented; Section \ref{sec:results} identifies the contribution of the dispersive fluxes and evaluates the work hypothesis indicated above. Finally, an extended  discussion of the results and implications for ASL parameterization is provided in Section \ref{sec:discussion}, with conclusions in Section \ref{sec:conclusions}.

\section{Dispersive Fluxes}\label{sec:theory}

When the temporally averaged equations of motion are spatially averaged, dispersive fluxes appear as an additional term that accounts for the effect of the spatial fluctuations of the time averaged variables. As a result, dispersive fluxes quantify the spatial correlations of temporally averaged quantities. Initially introduced by \cite{Wilson1977, Raupach1982}, dispersive fluxes can arise when a time averaged variable ($\overline{u}_i$, with the overline representing time averaging) is represented as the sum of a time and spatially averaged variable ($\langle\overline{u}_i\rangle$, with the $\langle \cdot \rangle$ representing spatial averaging) with a time averaged spatial deviation ($\overline{u}_i''$), for example, 
\begin{equation}
    \overline{u}_i = \langle\overline{u}_i\rangle + \overline{u}_i''.
\end{equation}
Hence, by using this decomposition, which is the result of spatially averaging time averaged variables, new covariances arise. These are the result of the spatial correlation of quantities averaged in time but varying with position. For example, the case for the heat fluxes may be written as 
\begin{equation}
    \langle \overline{w}''\overline{\theta}''\rangle = \langle \overline{w} \,\overline{\theta}\rangle - \langle \overline{w}\rangle \langle \overline{\theta}\rangle.
\end{equation}

Initially applied in atmospheric flows for the study of vegetated canopies \citep{Wilson1977,Raupach1981,Raupach1982,Raupach1994,Finnigan2000}, dispersive fluxes have been shown to potentially play an important role in the description of spatially averaged flow statistics \citep{Poggi2004,Poggi2008}. Early work in this area indicated that dispersive fluxes are negligible in the mean flow outside canopies \citep{Raupach1986,Cheng2002a}. In contrast, \cite{Poggi2004} showed, using a flume study, that the contribution to the momentum fluxes of the dispersive fluxes could be larger than 10\% in sparse canopies. Smaller dispersive fluxes of about 6\% were also measured by \cite{Mignot2009} in flow over a gravel bed. However, \cite{Bailey13} showed that in sparse, row-oriented canopies, the dispersive flux was over 20 \% of the magnitude of the turbulent flux. Similarly, a recent study demonstrated that dispersive fluxes are generated around canopy edges and that they can be large in the entry region of the canopy \citep{Moltchanov2015}. In addition, in urban-canopy studies, the contribution of dispersive fluxes has also been shown to be non-negligible \citep{Martilli2007}. For example, in LES studies of flow over random urban-like obstacles, the magnitude of the dispersive flux represents 15\% of the magnitude of the turbulent flux, and the peak of the dispersive flux corresponds to the top of the urban-like canopy \citep{Xie2008}. These results have been confirmed in flow over realistic urban surface by \cite{Giometto2016}. Similarly, studies of flow inside wind farms have shown that dispersive fluxes constitute a non-negligible fraction of the total vertical momentum flux \citep{Calaf2010}.

In this work, the concept of dispersive fluxes is extended and applied to the sensible heat flux with a double purpose. First as a means to account for the missing contribution to the surface energy budget resulting from the permanent spatial thermal heterogeneities present in the time averaged velocity and temperature fields. This idea is inline with the work of \cite{Zhouetal2018}. Second, as a potential means to develop new parameterizations for NWP models that inherently account for the effect of the unresolved thermal heterogeneities.

\section{Numerical Simulations and study cases}
\label{sec:Simul_cases}
In this section, the LES approach used for this study is briefly presented -- an overview of the LES method can be found in \cite{Moeng2015}, and more details on the procedure used here can also be found in \cite{BouZeid2004,BouZeid2005}, \cite{Calaf2011}, and \cite{Margairaz2018}. In the second part of this section details of the different study cases are presented. 

\subsection{Large-Eddy simulations framework}
\label{ssec:LES}
In LES, the time- and space-evolution of the turbulent flow is separated into resolved and filtered (modelled) scales. The resolved flow is determined by integrating the filtered incompressible Navier-Stokes (NS) equations  written in rotational form to ensure conservation of mass of the inertial terms \citep{Kravchenko1997}, and coupled to the advection-diffusion equation of heat using the Boussinesq approximation.  The dimensional form of the governing equations is therefore
\begin{equation}\label{eq:mass}
\partial_i{\tilde{u}_i}=0,
\end{equation}
\begin{equation}\label{eq:mom}
\partial_t{\tilde{u}_i} + \tilde{u}_j\left(\partial_j\tilde{u}_i-\partial_i\tilde{u}_j \right) = -\partial_i{\tilde{p*}} - \partial_j{\tau_{ij}^{\Delta}} + g\left(\frac{\tilde{\theta} - \langle\tilde{\theta}\rangle_{xy}}{ \langle\tilde{\theta}\rangle_{xy}}\right)\delta_{i3} + \tilde{f}_i,
\end{equation}\label{eq:heat}
\begin{equation}
\partial_t{\tilde{\theta}_i} + \tilde{u}_j\partial_j \tilde{\theta}=\partial_j\pi^{\Delta}_j.
\end{equation}
In these equations, $\tilde{u}_i$ ($i=1,2,3$) refer to the filtered velocity components in the three Cartesian directions (horizontal: $x,y$, and vertical: $z$), $\tilde{\theta}$ represents the filtered potential temperature, and $\tilde{p}^*$ denotes the dynamic modified pressure field. This is defined as $\tilde{p}^* = \tilde{p} + \frac{1}{3}\rho_0\tau_{kk}^{\Delta}  + \frac{1}{2} \tilde{u}_j\tilde{u}_j$, where the first term is the kinematic pressure, the second term is the trace of the sub-grid scale (SGS) stress tensor and the last term comes from the rotational form of the convective term. In equation \eqref{eq:mom}, the coupling term $g\left(\frac{\tilde{\theta} - \langle\tilde{\theta}\rangle_{xy}}{ \langle\tilde{\theta}\rangle_{xy}}\right)\delta_{i3}$ results from the Boussinesq approximation where $\langle\,\rangle_{xy}$ represents a horizontal average, and $\delta_{ij}$ is the Kronecker-delta operator. The flow is driven by a geostrophic forcing, imposed using the body force term $\tilde{f}_i=(\tilde{u}_2 -V_G)\nu_G\delta_{i1} - (\tilde{u}_1 -U_G)\nu_G\delta_{i2}$, where $(U_G,V_G)$ are the horizontal geostrophic velocity components, and $\nu_G=1.0\cdot10^{-4}$ Hz is the geostrophic frequency at a latitude of $43.3^{\circ}$ N. 

The deviatoric part of the SGS stress tensor is written using the eddy-viscosity approach and reads as $\tau_{ij}^{\Delta,d} = \tau_{ij}^{\Delta} - \frac{1}{3}\tau_{kk}^{\Delta}\delta_{ij} = 2\nu_{T}\tilde{S}_{ij}$, where $\nu_{T} = (C_S\Delta)^2|\tilde{S}|$ is the turbulent eddy-viscosity,  $\tilde{S}_{ij} = \frac{1}{2}\big(\partial_j \tilde{u}_i + \partial_i\tilde{u}_j\big)$ is the resolved strain rate tensor, and $C_S$ is the Smagorinsky coefficient \citep{Smagorinsky1963,Lilly1967}. This coefficient is computed dynamically using the Lagrangian scale-dependent dynamic model of \cite{BouZeid2005}. Similarly for temperature, the SGS temperature diffusion is given by $\pi^{\Delta}_j = -\nu_T/Pr_{sgs}\partial_j\tilde{\theta} = - (D_S\Delta)^2|\tilde{S}|\partial_j\tilde{\theta}$, where the coefficient $D_S$ is computed dynamically using a Lagrangian scale-dependent dynamic model for scalars \citep{Calaf2011}. 

The numerical implementation is based on \cite{Albertson1999b}, later modified by \cite{BouZeid2005}, \cite{Calaf2011}, and \cite{Margairaz2018}. This pseudo-spectra code treats the horizontal derivatives in Fourier space, the vertical derivatives are computed using second order finite differences on the vertically staggered grid, and the second order Adam-Bashforth scheme is used for time integration. The lateral boundary conditions are as a result periodic. The top boundary conditions are prescribed using a stress-free lid condition for the horizontal velocity ($\partial_z\tilde{u}_i=0, i=1,2$) and a constant flux for the temperature ($\partial_z\tilde{\theta}=cst$). The  non-penetration condition ($\tilde{w}=0$) is imposed on the vertical velocity at the top and bottom of the domain. Monin-Obukhov similarity theory (MOST) \citep{MoninObukhov54} is used for the bottom boundary condition for the horizontal velocity and temperature. The latter gives a formulation for the surface shear-stress and vertical heat flux. The drag from the underlying surface is entirely modeled through the equilibrium logarithmic law for rough surfaces \citep{Karman1931,Prandtl1932}. The surface friction velocity $u_*$, related to the shear stress, is given by
\begin{equation}\label{eq:wallstress}
u_*^2 = \left[\frac{\kappa}{\ln\left(\frac{\Delta z/2}{z_0}\right)+ \psi_m\left(\frac{\Delta z/2}{L}\right)}\right]^2 \Big(\hat{\tilde{u}}_1^2(x,y,\Delta z/2) + \hat{\tilde{u}}_2^2(x,y,\Delta z/2) \Big).
\end{equation} 
In this equation, $\hat{\tilde u}_i$ is the velocity field filtered at $2\Delta_{LES}$ and sampled at $\Delta z/2$ where $\Delta z$ is the vertical grid size $\Delta_{LES}=\sqrt{\Delta x \Delta y}$. The aerodynamic roughness length is denoted by $z_0$ and $\kappa=0.4$ is the von K\'arm\'an constant. The stability correction function of momentum $\psi_m$ is computed using Brutsaert's formulation \citep{Brutsaert2005} and depends on atmospheric stability assessed by the local Obukhov length $L=-\frac{u_*^3 \theta_S}{\kappa g q_s}$. Here, $\theta_S$ is the surface temperature, $g$ is the gravitational acceleration, and $q_s$  denotes the surface heat flux \citep{Brutsaert1982}. The wall shear-stress is then dynamically projected over the horizontal directions using the unit direction vector of the horizontal velocity sampled at $z=\Delta z/2$ and filtered at $2\Delta_{LES}$ \citep{BouZeid2004,Hultmark2013}.  

Similarly, the vertical sensible heat flux is computed as
\begin{equation}\label{eq:qsfc}
q_s=\frac{\left[\theta_s(x,y) - \tilde{\theta}(x,y,\Delta z/2)\right] }{\left[\ln\left(\frac{\Delta z/2}{z_{0s}}\right) +\psi_s\left(\frac{\Delta z/2}{ L}\right)\right]}u_*
\end{equation}
where $\psi_s$ is the stability correction function for the temperature. The corresponding vertical derivatives of the horizontal velocities and temperature are imposed at the first grid point of the vertically staggered grid following \cite{Albertson1999b}. This framework was developed for idealized homogeneous surfaces. The authors are aware of the limitations of the surface parametrization as mentioned in Basu's cautionary note \citep{Basu2017}. However, it is also known that this approach provides acceptable results in the case of non-homogeneous conditions \citep{BouZeid2004,Stoll2006a}. 

\subsection{Study cases}
\label{sec:cases}

\begin{figure}[t]
    \centering
    \includegraphics[width=\textwidth]{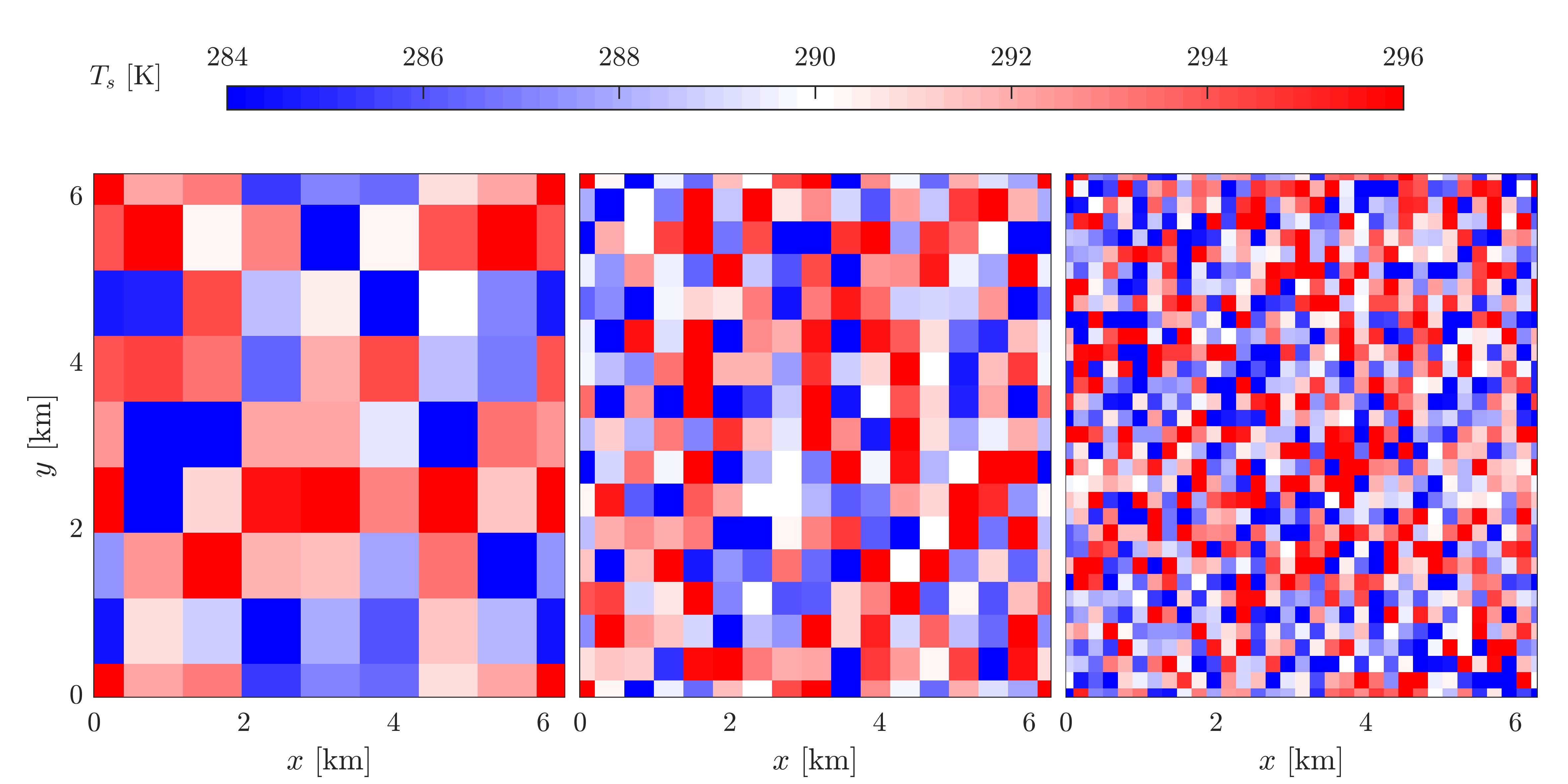}
    \caption{Surface temperature distribution for the patches cases. From left to right: 800 m patches (HT-X-s800.$v_1$), 400 m patches (HT-X-s400), and 200 m patches (HT-X-s200), }
    \label{fig:surface_temp}
\end{figure}

In order to understand the effect of a heterogeneous surface compared to a homogeneous surface, the simulations are separated in two sets. In the first set, a total of seven configurations are considered, all with a homogeneous surface temperature fixed at a value of $\theta_S=290$ K, and for which the geostrophic wind speed has been increased from 1 to 15 m/s (\textit{i.e.} $U_g =$ 1, 2, 3, 4, 6, 9, 15 m/s). These cases are referred hereafter as HM-X, where X indicates the geostrophic wind speed corresponding case (see Table \ref{table:runsummary}). In the second set, the surface temperature is distributed in patches, where the temperature of the patches is determined using a Gaussian distribution with a mean temperature of $290$ K and a standard deviation of $5$ K. In this case, three different patch sizes have been considered (\textit{i.e.} $l_h = $ $800$, $400$, and $200$ m). The sizes of the heterogeneities have been chosen to be of similar size ($l_h/l_e \sim 1$), half the size ($l_h/l_e \sim 1/2$), and about a quarter of the size ($l_h/l_e \sim 1/4$) of to the most energetic eddies within the represented thermal boundary layer, assuming that the integral length scale in the ABL is of the order of the BL height ($l_h \sim l_e \sim z_i$). Further, these heterogeneities are typically not resolved in NWP models. These cases have also been studied for the different geostrophic winds indicated above. These cases are referred hereafter as HT-X-sYYY, where X indicates the corresponding geostrophic wind speed, and $sYYY$ refers to the size of the patches (\textit{e.g.}, HT-1-s800 would be the heterogeneous case with patches of $800$ m, and forced with $U_g = 1$ m/s). Additionally, for the case with larger patches, three different random distributions of the patches have been considered to evaluate the potential effect of a given surface distribution for all geostrophic wind speeds. This is further indicated with the indicator $v_i$ with $i = 1,2,3$. Therefore, in this second set of study cases a total of 35 different configurations have been considered (see Table \ref{table:runsummary}). Figure \ref{fig:surface_temp} shows the surface temperature distributions used for the different heterogeneous surface conditions. These temperature distributions emulate the surface conditions observed in \cite{Morrison2017}, where measurements of the surface temperature were taken with a thermal camera at the SLTEST site of the US ARMY Dugway Proving Ground in Utah, USA. This is an ideal site with uniform roughness and a large unperturbed fetch; where surface thermal heterogeneities are naturally created by differences in surface salinity. 

In all studied cases the surface roughness is kept constant at $z_0=0.1$ m, and the initial boundary layer height is set to $z_i=1000$ m. The temperature profile is initialized with a mean air temperature of $285$ K. At the top of the initial boundary layer, a capping inversion of $1000$ m is used to limit its growth. The strength of this inversion is fixed at $\Gamma = 0.012$ K/m. The ABL is considered dry and the latent heat flux is neglected in all the cases. Further, in all simulations, the surface heat flux is computed using MOST as explained in section \ref{ssec:LES}. Hence, to ensure a degree of homogeneity within each patch and a certain degree of validity of MOST, note that even for the heterogeneous cases with the fewest amount of grid points per patch, a minimum of eight grid points is granted in each horizontal direction. The domain size is set to $(L_x,L_y,L_z)=(2\pi,2\pi,2)$ km at a grid size of  $(N_x,N_y,N_z)=(256,256,256)$ resulting in an horizontal resolution of $\Delta x = \Delta y = 24.5$ m and a vertical grid spacing of $\Delta z = 7.8$ m. A time step of $\Delta t = 0.1$ s is used to ensure the stability of the time integration. This setup is very similar to the one used by \cite{Salesky2016}. 

The two sets of simulations are run spanning a large range of geostrophic forcing conditions. This allows us to study its effect on the structure of the CBL above a patchy surface compared to a homogeneous surface. The range of $U_g$ covers values between $1$ m/s and $15$ m/s. The procedure used to spin up the simulations is the following: a spin-up phase of four hours of real time is used to achieve converged turbulent statistics, which is then followed by an evaluation phase. During the latter, running averages are computed for the next hour of real time.  Statistics have been computed for averaging times of five minutes to one hour, showing statistical convergence at $30$ min averages with negligible changes between the $30$ min and the one hour averages. 

Table \ref{table:runsummary} presents a summary of the simulation statistics for the homogeneous and heterogeneous surface cases. The values of the Obukhov length $L=-\frac{u_*^3 \theta_S}{\kappa g \overline{(w'\theta')_S}}$, the stability parameter $-z_i/L$, the convective velocity $w_*=\left[\frac{g}{\bar \theta_S}z_i\overline{(w'\theta')_S}\right]^{\frac{1}{3}}$, and the temperature scale $\theta_*=\left[\frac{g}{\bar \theta_S}z_i\right]^{-\frac{1}{3}}\left[\overline{(w'\theta')_S}\right]^{\frac{2}{3}}$ have been obtained using the planar averages of the thirty minutes averages of the friction velocity $u_*$ (from equation \ref{eq:wallstress}), the sensible heat flux $\overline{(w'\theta')_S}$ (from equation \ref{eq:qsfc}) and the height of the boundary layer $z_i$. The simulations cover a wide range of atmospheric stability regimes ranging from $-z_i/L < 5$ to $-z_i/L > 700$, and hence spanning from near neutral to highly convective scenarios. 

\begin{table}[t]\begin{center}
    \begin{tabular}{lrrrrrrrrr}
    \hline
    Name &$U_g$ & $z_i$ & $L$ & $-z_i/L$ & $u_*$ & $\theta_*$ & $\langle\overline{(w'\theta')_S}\rangle$ & $w_*$ & $w_*/u_*$ \\
    
     & [m/s] & [m] & [m] & [-] & [m/s] & [K] & [mK/s] & [m/s] & [-]\\[.3em]
    \hline
    HM-1&$1$&$1312.5$&$-3.95$&$331.95$&$0.17$&$0.52$&$0.09$&$1.44$&$8.58$\\
    HM-2&$2$&$1304.7$&$-5.84$&$223.30$&$0.19$&$0.45$&$0.09$&$1.43$&$7.54$\\
    HM-3&$3$&$1289.1$&$-11.27$&$114.40$&$0.23$&$0.36$&$0.08$&$1.42$&$6.05$\\
    HM-4&$4$&$1281.2$&$-19.81$&$64.67$&$0.28$&$0.29$&$0.08$&$1.41$&$5.02$\\
    HM-6&$6$&$1273.4$&$-44.39$&$28.69$&$0.37$&$0.23$&$0.08$&$1.42$&$3.83$\\
    HM-9&$9$&$1296.9$&$-96.79$&$13.40$&$0.49$&$0.18$&$0.09$&$1.44$&$2.96$\\
    HM-15&$15$&$1328.1$&$-266.33$&$4.99$&$0.71$&$0.14$&$0.10$&$1.49$&$2.11$\\[.3em]
    \hline\hline\\[-.8em]
    HT-1-s800.$v_1$&$1$&$1421.9$&$-1.99$&$715.47$&$0.16$&$0.97$&$0.16$&$1.75$&$10.80$\\
    HT-2-s800.$v_1$&$2$&$1382.8$&$-3.21$&$430.72$&$0.19$&$0.81$&$0.15$&$1.73$&$9.20$\\
    HT-3-s800.$v_1$&$3$&$1359.4$&$-5.85$&$232.20$&$0.23$&$0.64$&$0.14$&$1.70$&$7.53$\\
    HT-4-s800.$v_1$&$4$&$1343.8$&$-10.09$&$133.23$&$0.27$&$0.51$&$0.14$&$1.67$&$6.28$\\
    HT-6-s800.$v_1$&$6$&$1335.9$&$-23.80$&$56.14$&$0.35$&$0.37$&$0.13$&$1.63$&$4.72$\\
    HT-9-s800.$v_1$&$9$&$1343.8$&$-67.09$&$20.03$&$0.49$&$0.26$&$0.13$&$1.62$&$3.34$\\
    HT-15-s800.$v_1$&$15$&$1359.4$&$-195.56$&$6.95$&$0.69$&$0.18$&$0.13$&$1.62$&$2.34$\\
    \hline\\
    HT-1-s400&$1$&$1486.6$&$-0.43$&$3432.62$&$0.10$&$1.57$&$0.15$&$1.01$&$10.52$\\
    HT-2-s400&$2$&$1463.2$&$-1.65$&$897.02$&$0.15$&$1.06$&$0.16$&$0.99$&$6.47$\\
    HT-3-s400&$3$&$1452.0$&$-3.66$&$397.08$&$0.20$&$0.79$&$0.16$&$0.94$&$4.76$\\
    HT-4-s400&$4$&$1434.2$&$-8.42$&$170.45$&$0.26$&$0.58$&$0.15$&$0.90$&$3.50$\\
    HT-6-s400&$6$&$1406.2$&$-22.48$&$62.57$&$0.35$&$0.40$&$0.14$&$0.88$&$2.51$\\
    HT-9-s400&$9$&$1389.5$&$-62.47$&$22.26$&$0.48$&$0.27$&$0.13$&$0.83$&$1.73$\\
    HT-15-s400&$15$&$1404.0$&$-197.22$&$7.13$&$0.70$&$0.18$&$0.13$&$0.76$&$1.08$\\
    \hline\\
    HT-1-s200&$1$&$1438.6$&$-0.83$&$1741.47$&$0.12$&$1.28$&$0.15$&$0.93$&$7.74$\\
    HT-2-s200&$2$&$1434.2$&$-1.54$&$932.71$&$0.15$&$1.04$&$0.15$&$0.89$&$6.02$\\
    HT-3-s200&$3$&$1429.7$&$-3.79$&$377.06$&$0.20$&$0.77$&$0.15$&$0.83$&$4.17$\\
    HT-4-s200&$4$&$1434.2$&$-7.97$&$180.07$&$0.25$&$0.59$&$0.15$&$0.79$&$3.12$\\
    HT-6-s200&$6$&$1401.8$&$-23.00$&$61.04$&$0.35$&$0.40$&$0.14$&$0.75$&$2.13$\\
    HT-9-s200&$9$&$1373.9$&$-64.68$&$21.26$&$0.48$&$0.26$&$0.13$&$0.70$&$1.45$\\
    HT-15-s200&$15$&$1382.8$&$-206.70$&$6.70$&$0.70$&$0.18$&$0.12$&$0.63$&$0.89$\\
    \hline\\
    \end{tabular}
    \caption{Summary of the study cases and the corresponding most relevant simulation statistics for 30 minutes averages. The homogeneous cases are referenced as HM, and cases with heterogeneous surfaces are referred as HT. In the case of the 800 m patches, only the statistics for HT-X-s800.$v_1$ are presented because the values for cases $v_2$ and $v_3$ are very similar. The proprieties presented in the table are the geostrophic forcing velocity ($U_g$), the boundary layer height ($z_i$), the Obukhov length ($L$), the stability factor ($-z_i/L$), the friction velocity ($u_*$), the temperature scale $\theta_*$, the planar averaged surface heat flux $\langle\overline{(w'\theta')_S}\rangle$), and the convective velocity scale ($w_*$).}
    \label{table:runsummary}
\end{center}\end{table}

\section{Result}
\label{sec:results}
\subsection{Impact of the heterogeneity on the ABL scale structures}\label{ssec:ABLstruct}

\begin{figure}[t]
    \centering
    \includegraphics[width=1\textwidth]{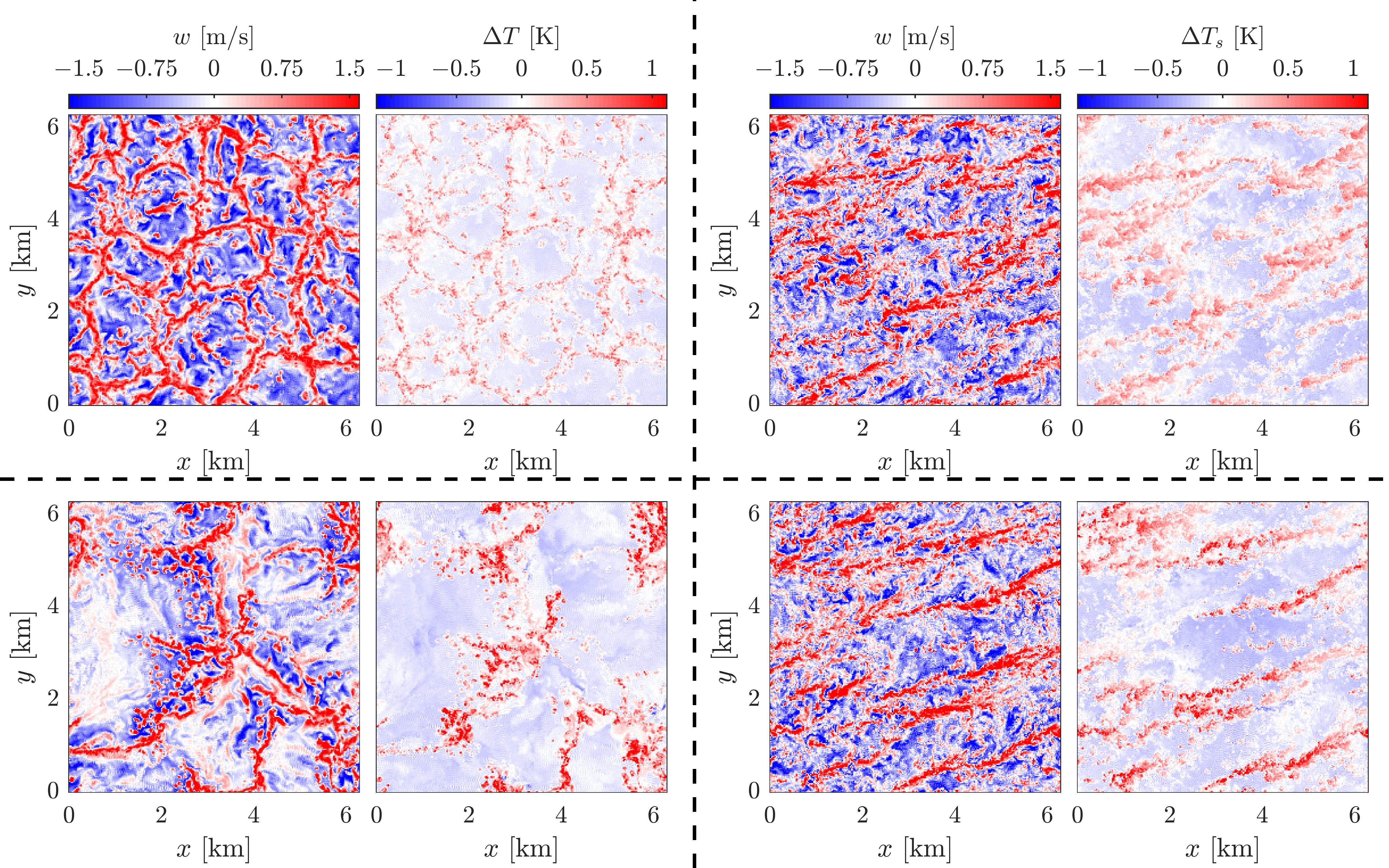}
    \caption{Instantaneous 2D-horizontal ($x-y$) fields at $z = 100$ m and $t=4$ hours for the vertical velocity $w=w(x,y,z,t)$ and temperature difference $\Delta T = T(x,y,z,t)=\langle T \rangle_{x,y}(z,t)$. The top four subfigures correspond to cases over a homogeneous surface with a geostrophic wind of $U_g = 1$ m/s (HM-1) for the two most left subfigures and of $U_g = 15$ m/s for the two most right subfigures (HM-15). The bottom four subfigures correspond to the corresponding heterogeneous cases with patches of size 800 m (HT-1-s800.$v_1$ \& HT-15-s800.$v_1$).}
    \label{fig:InstFields}
\end{figure}

As a first step of the analysis, we are interested in investigating the effect of the surface thermal heterogeneities on the flow field, as well as the effect of increasing the geostrophic forcing. For this purpose, Figure \ref{fig:InstFields} presents instantaneous fields of the vertical velocity and temperature difference at a height of 100 m ($\sim 1/10 \,z_i$). The instantaneous snapshots for the homogeneous cases (top row plots) are good for understanding the effect of increasing geostrophic forcing. In the case corresponding to a low wind ($U_g = 1$ m/s, top left), the expected convective cells can be observed, both in the velocity and temperature fields. These convective open cells have a diameter of 2 to 4 km, corresponding to observations made in the CBL \citep{Konrad1970,Weckwerth1990,Bennett2010}. As the geostrophic wind is increased ($U_g = 15$ m/s), it is very interesting to see the transformation of the flow structure, from a convective cell geometry into a roll structure. The transition between cell and roll structure is related to the increase in shear stress, which destroys the original convective cell structure. In order to quantify this transition, \cite{Salesky2016} developed a metric based on the two-point correlation function of the vertical velocity in cylindrical coordinates $R_{ww}(r_{\eta},r_{\phi},z)$. This method detects the angular dependencies present in the roll-type convection using the statistical range of the two-point correlation function. The statistical range, defined as
\begin{equation}
R(r_{\eta})=\max_{r_{\phi}}\big[R_{ww}(r_{\eta},r_{\phi},z)\big] - \min_{r_{\phi}}\big[R_{ww}(r_{\eta},r_{\phi},z)\big], 
\end{equation}
will be large for the roll-type convection and small for cell-type convection. Following this observation, the Roll Factor can be defined as 
\begin{equation}
\mathcal{R}=\max_{r_{\eta}}\big[R(r_{\eta})|r_{\eta}/z_i \le 0.5 \big].
\end{equation}
In this analysis, the radial lag is cut at $z/z_i=0.5$ because only large convection structures are of interest. Using this metric the study cases presented in Figure \ref{fig:InstFields} have a Roll Factor of $\mathcal{R}\sim 0.1$ for $U_g=1$ m/s and $\mathcal{R}\sim 0.3$ for $U_g=15$ m/s, matching the conceptual structure defined in \cite{Salesky2016} well. 

Interestingly, the original convective cell structure is also modified when the flow is overlaid on a thermally heterogeneous surface. In this case (bottom row left), the convective cell structure is adjusted to the surface thermal patchiness. The instantaneous snapshot of the flow field illustrates how cells merge into larger cells, or break into much smaller cells or updrafts that concentrate on the surroundings of the larger cells. Therefore, in this case the characteristic length scale of the larger convective cells is related to the length scale of the surface thermal heterogeneities ($l_h$) or larger, if there are nearby patches with similar surface temperature. For smaller updrafts, the characteristic length scale is of the order of 200 m to 400 m, compatible with the observations made by \cite{Bennett2010} over Oklahoma. This length scale is smaller than the individual surface temperature patch sizes ($l_h$). Interestingly, when the geostrophic wind increases ($U_g = 9$ m/s), the convective-cell structure is also destroyed, similar to the homogeneous case, along with the footprint of the surface patchiness (bottom row right). In this case, the Roll Factor is $>0.25$, very similar to the one obtained in the case of a homogeneous surface temperature. It is therefore clear, that under the effect of moderate geostrophic forcing the impact of the surface thermal heterogeneities is blended, and its corresponding impact reduced.

\begin{figure}[t]
    \centering
    \includegraphics[width=.95\textwidth]{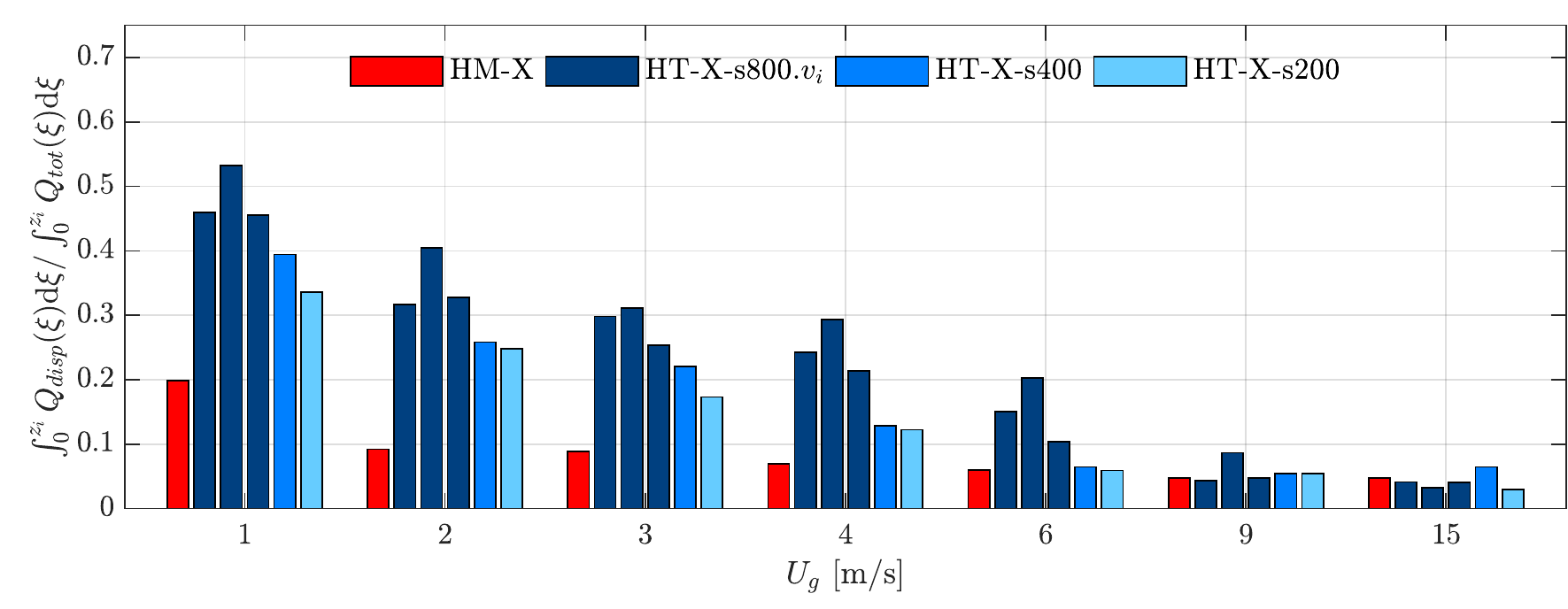}
    \caption{Integral fraction of sensible heat flux accounted by the dispersive fluxes as a function of geostrophic wind when averaged over a 30 min time period. (red) indicates the homogeneous cases; (dark-blue) heterogeneous cases with 800 m patches; (blue) heterogeneous cases with 400 m patches; (light-blue)  heterogeneous cases with 200 m patches.}
    \label{fig:DispFluxes}
\end{figure}

\subsection{Dispersive fluxes}\label{ssec:dispflux}

This progressive blending of the thermal heterogeneities with the increasing geostrophic wind can be better observed through the magnitude of the dispersive fluxes computed for the study cases introduced in section \ref{sec:cases}. To quantify the weight of the dispersive flux in the calculation of the total sensible heat flux, the following metric is introduced,  \begin{equation}
    \dfrac{\int_{0}^{z_i}Q_{disp} (\xi)\,\mathrm{d}\xi}{\int_{0}^{z_i}\Big[Q_{Reynolds} (\xi) + Q_{SGS} (\xi) + Q_{disp} (\xi)\Big]\,\mathrm{d}\xi} = \dfrac{\int_{0}^{z_i}Q_{disp} (\xi)\,\mathrm{d}\xi}{\int_{0}^{z_i} Q_{tot} (\xi)\,\mathrm{d}\xi}.
\end{equation}
In the equation above, $Q_{disp}$ represents the dispersive flux, $Q_{Reynolds}$ is the planar-averaged resolved sensible heat flux, and $Q_{SGS}$ is the planar-averaged SGS contribution of the sensible heat flux. Because the fluxes are integrated over the full ABL column, this metric represents the integral fraction of sensible heat flux that arise from the dispersive fluxes. Note that the boundary layer height is taken as the height where the sensible heat flux crosses the zero value before the capping inversion. Figure \ref{fig:DispFluxes} illustrates these results for values of fluxes computed for a 30 minute time period. From this figure, it is very interesting to see the progressive decrease in the contribution of dispersive fluxes with increasing geostrophic wind speed, which is a result of the increased blending as illustrated above, up to a point in which the effect of the surface thermal heterogeneities does not have an impact on the dispersive fluxes. These results seem therefore to indicate the existence of two differentiated regimes: one in which the contribution of the dispersive fluxes is intrinsic to the surface thermal heterogeneities, and hence represent a measure of the impact of the surface on the flow; and a second regime in which the dispersive fluxes are fully due to the turbulence coherent structure, or consistency in time related to the surface induced shear. In the case where the dispersive fluxes are related to the surface thermal heterogeneities it is very interesting to see how these can account for more than 40\% of the total sensible heat flux for the cases HT-1-s800.$v_i$ ($i=1,2,3$), about 40\% for HT-1-s400, and about 33\% for the case HT-1-s200 when integrated over the full ABL depth. Also, from the results presented in Figure \ref{fig:DispFluxes}, it can be observed that different surface spatial distributions of equal size ($l_h$) and equal standard deviation (cases HT-1-s800.$v_i$ with $i=1,2,3$) lead to very similar values of the dispersive flux contribution. This is an important result because it illustrates the independence of the results here extracted from any specific spatial distribution of the thermal heterogeneities.

\begin{figure}[t]
    \centering
    \includegraphics[width=1\textwidth]{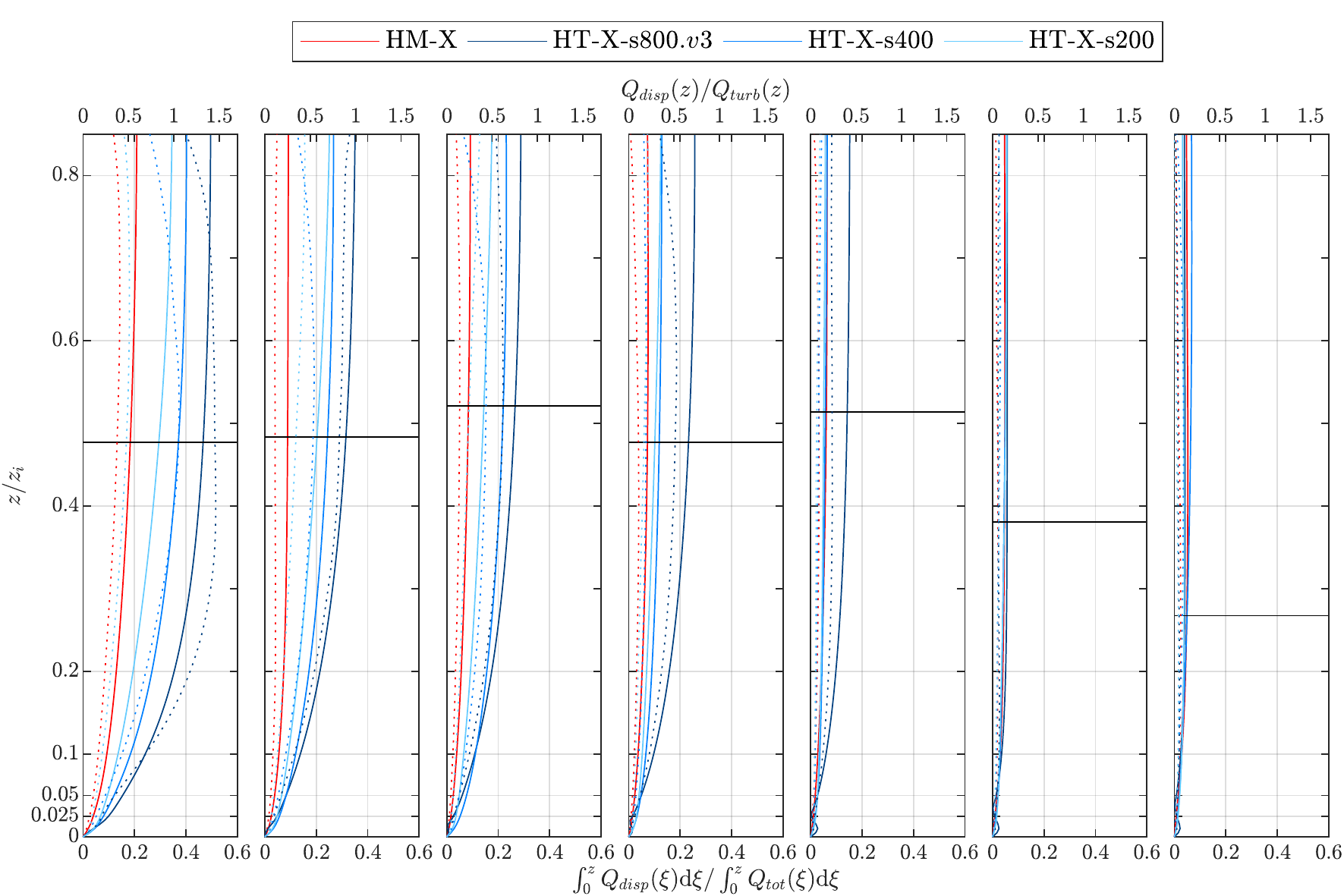}
    \caption{Continuous lines represent the vertical profiles of the integral fraction of sensible heat flux accounted by the dispersive fluxes for 30 min averages and as a function of geostrophic wind speed (from left to right: $U_g=1,\,2,\,3,\,4,\,6,\,9,\,15$ m/s). The horizontal black line illustrates the height above which the contribution of the dispersive fluxes does not change by more than 10\%. The doted-lines illustrate the local ratio between the dispersive flux and the total sensible heat flux at a give height $z$.}
    \label{fig:DispFluxes_Profiles}
\end{figure}

Because, the vertically integrated measures of the dispersive fluxes illustrated in Figure \ref{fig:DispFluxes} do not provide any information on the prevalence of the dispersive fluxes as a function of height, Figure \ref{fig:DispFluxes_Profiles} analyzes this next. In this case, Figure \ref{fig:DispFluxes_Profiles} illustrates vertical profiles of the integral fraction of the dispersive flux with respect to the total sensible heat flux for 30 min time averages as a function of height, 
\begin{equation}
    \frac{\int_{0}^{z}Q_{disp} (\xi)\,\mathrm{d}\xi}{\int_{0}^{z}\Big[Q_{Reynolds} (\xi) + Q_{SGS} (\xi) + Q_{disp} (\xi)\Big]\,\mathrm{d}\xi} = \dfrac{\int_{0}^{z}Q_{disp} (\xi)\,\mathrm{d}\xi}{\int_{0}^{z} Q_{tot} (\xi)\,\mathrm{d}\xi}.
\end{equation}
Note that in this case, both the numerator and the denominator represent the integral up to a given height $z$. The ratio of the dispersive flux over the total sensible heat flux represents the averaged cumulative contribution of the dispersive flux up to a given height. It is for this reason for example that in all subplots the dispersive flux contribution is zero at the surface. Moving away from the surface, the contribution of the dispersive fluxes increases quickly with height until reaching saturation at a height around $z/z_i \sim 0.3 - 0.5$, depending on the geostrophic wind. In this figure, the horizontal black line illustrates the height above which the contribution of the dispersive fluxes does not change by more than 10\%. From the profiles, it is also interesting to see that close to the surface ($z/z_i \sim 0.02$) the contribution of the dispersive fluxes can be as much as 5 -- 10\% for a spatial average spanning through the full domain. Similar to what had been observed in Figure \ref{fig:DispFluxes}, the difference in net contribution to the total sensible heat flux by the dispersive fluxes diminishes with increasing geostrophic forcing, up to the point that there is no difference between the homogeneous and heterogeneous cases (overlap between all vertical profiles). This result is also inline with the earlier hypothesis regarding the existence of two regimes, one in which the dispersive fluxes are directly correlated with the surface thermal heterogeneities, and another one in which they are related to the surface shear induced turbulence structure. Also note that overlaid in Figure \ref{fig:DispFluxes_Profiles} there is represented the ratio of the contribution of the dispersive fluxes versus that of the turbulent contribution (dotted lines). This complementary illustration of the results further shows how in the low geostrophic wind cases dispersive fluxes are relevant, potentially being of equal or larger value than the turbulent fluxes. Interestingly, even close to the surface ($z/z_i \sim 0.02$) the contribution of dispersive fluxes is non-negligible.

\begin{figure}[t]
    \centering
    \includegraphics[width=.8\textwidth]{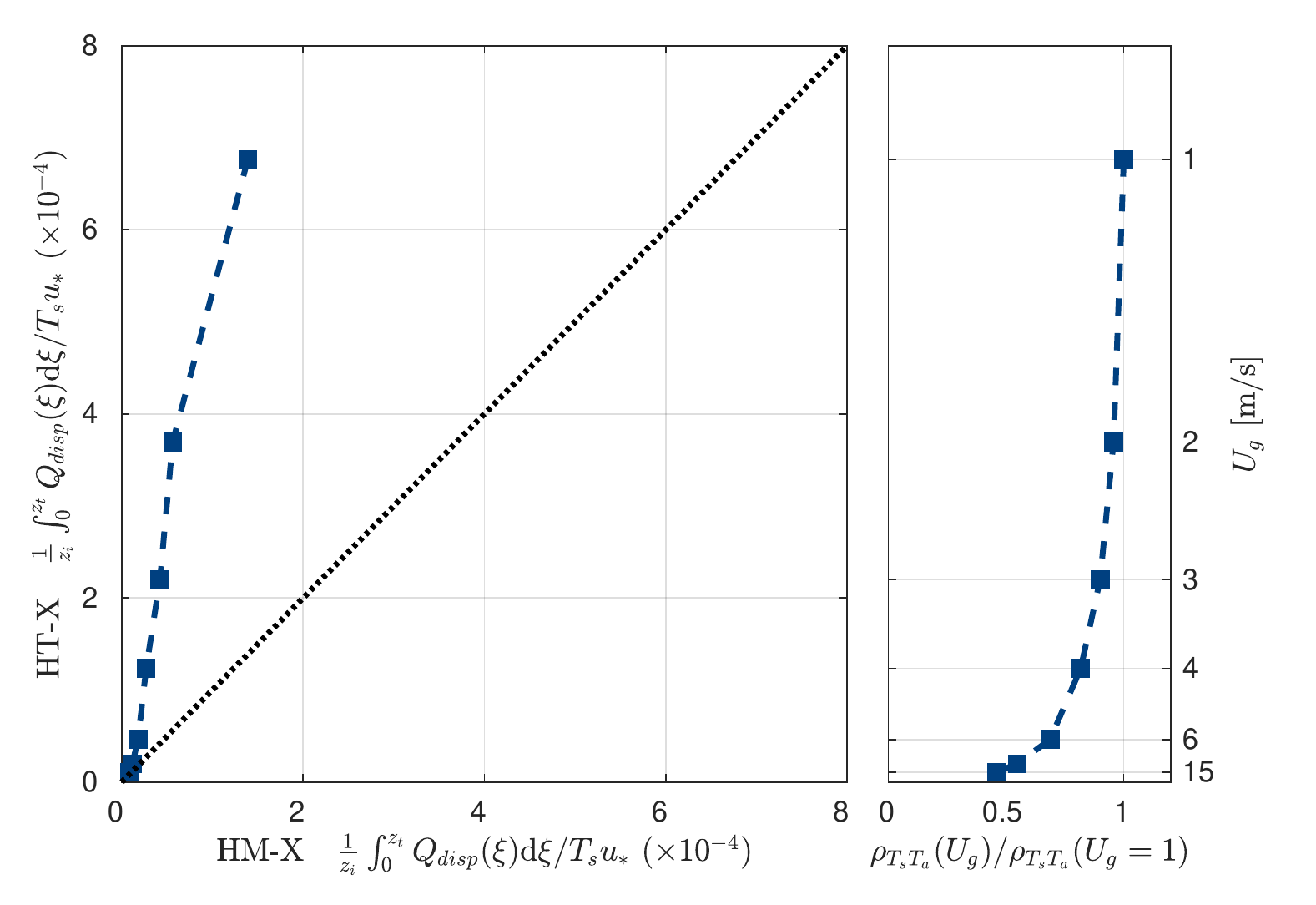}
    \caption{Left: Correlation in the contribution of the dispersive fluxes over the full ABL between the cases with a heterogeneous thermal surface (averaged over the different patch sizes) and the case with a homogeneous thermal surface. Right: Correlation between the surface temperature and air temperature at $z/z_i = 0.05$ as a function of the geostrophic wind.}
    \label{fig:FluxCorr}
\end{figure}

To further explore this direct relationship between the dispersive fluxes and the surface imposed thermal heterogeneities, Figure \ref{fig:FluxCorr} presents a correlation analysis that quantifies the relationship between the surface temperature distribution and the corresponding one at $z/z_i = 0.1$ (right subplot). It can be observed that the correlation is maximum in the case with the weakest geostrophic wind, and that this decreases with increasing wind. Interestingly, the cases presenting the largest correlations between the surface and air temperature correspond well with the cases in which the integral contribution of the dispersive fluxes is much larger than those found over the  homogeneous surfaces. This is well illustrated in Figure \ref{fig:FluxCorr} (right). This important relationship between the surface and air temperature at $z/z_i = 0.1$ is interpreted as responsible for the relevant contribution of the dispersive flux on the overall heat flux (see Figure \ref{fig:FluxCorr} right). Further, the authors believe that this result could be exploited to develop simplified parametrizations of the dispersive fluxes based on remote surface thermal measurements. In the case of the strongest geostrophic winds, this correlation is reduced to a minimum ($< 50\%$), and the dispersive fluxes match well with those measured over homogeneous surfaces indicating that they are unrelated to the surface thermal patchiness. 

\begin{figure}[t]
    \centering
    \includegraphics[width=.95\textwidth]{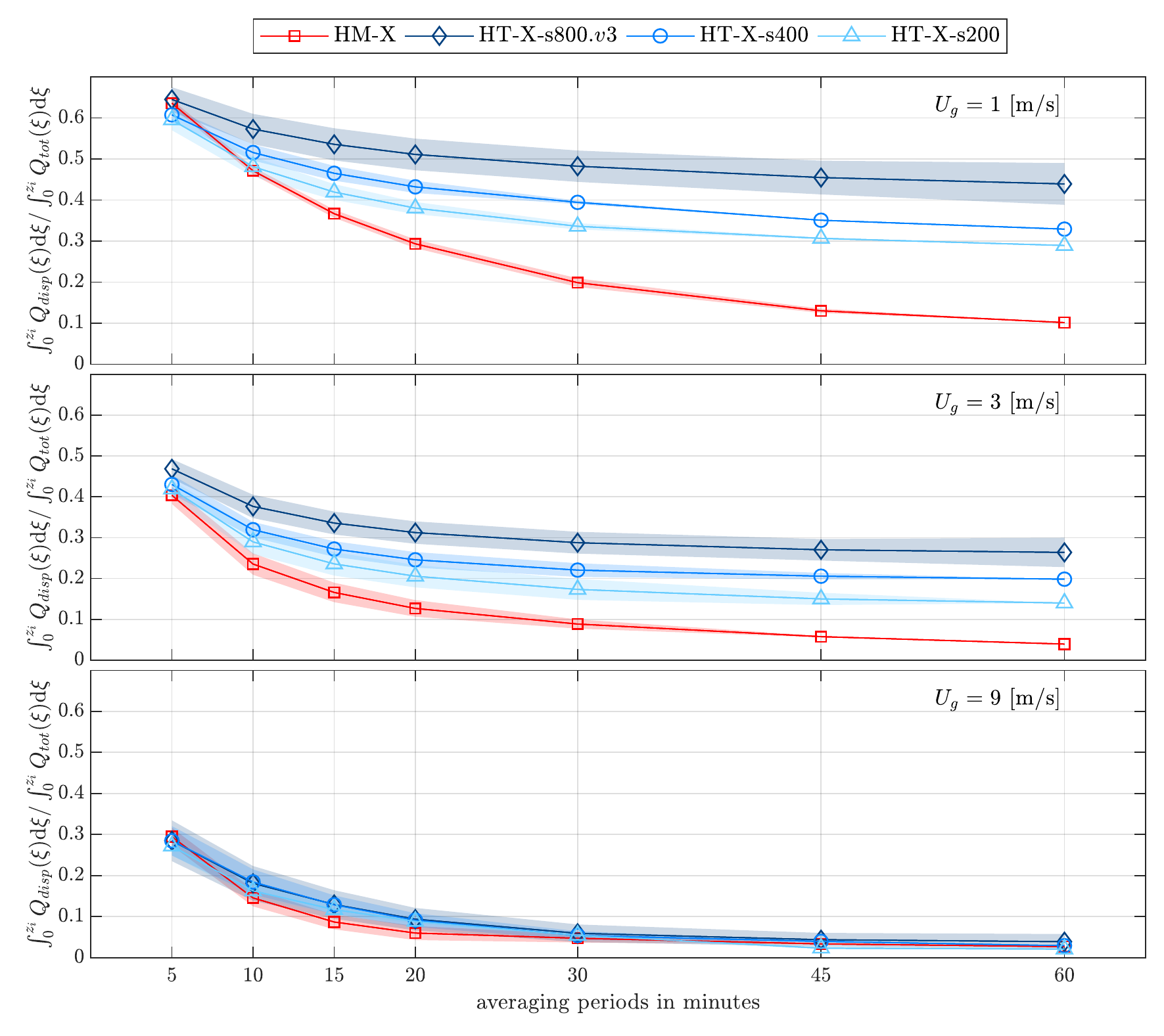}
    \caption{Integral fraction of sensible heat flux accounted by the dispersive fluxes as a function of averaging time, and geostrophic wind intensity (top $U_g = 1$m/s, middle $U_g = 3$m/s and bottom $U_g = 9$m/s) when averaged over a 30 min time period. (red) indicates the homogeneous cases; (dark-blue) heterogeneous cases with 800 m patches; (blue) heterogeneous cases with 400 m patches; (light-blue)  heterogeneous cases with 200 m patches.}
    \label{fig:TimeAvg}
\end{figure}

Finally, because the definition of dispersive fluxes is intrinsic, or fully dependent, on the definition of a specific time averaging operation as explained in Section \ref{sec:theory}, it is important to evaluate the dependence of the values obtained above as a function of the time averaging operation. Hence, Figure \ref{fig:TimeAvg} illustrates the integral fraction of sensible heat flux accounted by the dispersive fluxes as a function of time averaging, and geostrophic wind intensity. In both, top and middle subplots, in which the geostrophic forcing is weak and hence the contribution of the dispersive fluxes is related to the surface thermal heterogeneities, there is a maximum decrease in dispersive flux contribution with increasing time average from 5 minutes to 30 minutes ranging between 30\% -- 50\%. Note that this decrease is much smaller ($< 15\%$) beyond the 20-min average period,  and that the contribution of the the dispersive fluxes remains relevant even at 60 min averages, ranging between 30\% to 45\%
in very weak geostrophic winds ($U_g = 1$m/s), and between 10\% to 30\% in moderate geostrophic winds ($U_g = 3$m/s).

Through further analysis of the dispersive flux contribution in the homogeneous cases, it is also interesting to note that these are also important in either geostrophic forcing condition in short averaging times. Therefore, one can interpret the dispersive fluxes as being generated by the coherent spatial distribution of the turbulent flow in very short averaging times. As the averaging is increased, the dispersive fluxes contribution are a clear representation of the existence of surface heterogeneity and its effects on the flow (as also illustrated in Figure \ref{fig:FluxCorr}), and in the case the surface is homogeneous, or the geostrophic forcing is strong and blends the surface heterogeneity, then the dispersive fluxes partially account for the structure of the turbulent flow. This is also well illustrated by the bottom subplot in Figure \ref{fig:TimeAvg}, representative of the case where $U_g = 9$m/s and the effect of the surface heterogeneities is quickly blended. However, note that the value of the dispersive fluxes found on large averaging times over the homogeneous conditions are small, and if the LES domain used was larger, these would be even smaller and hence negligible. We believe that the residual value measured for the dispersive fluxes on the homogeneous cases results from the locking of large coherent structures induced by the limited size of the LES domain and the periodicity of the numerical algorithm as indicated in \citet{Munters2016}. Realize however, that in order to overcome this limitation the associated numerical cost would make it very difficult to consider the large amount of study cases presented here. Nevertheless, to ensure that the measurement of the dispersive fluxes over thermally heterogeneous surfaces represent well the flow physics and are not the result of numerical artifacts, a test simulation with a four times larger domain was also run for the HT-1-s800 study case. In this case, results corresponded well with those obtained with the smaller domain, indicating that in the heterogeneous cases the LES domain size did not affect the results presented.

\section{Dispersive fluxes in the surface layer}\label{sec:resultsII}

The results presented to this point have been horizontally averaged over the full domain. Therefore, while providing a measure of the overall contribution of the dispersive fluxes in the bulk ABL transport processes as a function of differential geostrophic and surface forcings, this approach does not provide an answer a-priory to the opening hypotheses of this work, namely: 1. Can dispersive fluxes provide a means of capturing the effect induced by the unresolved surface heterogeneities in near-future NWP resolutions ($\sim 100m$)? 2. Can dispersive fluxes explain the non-closure of the surface energy budget?

\begin{figure}[t]
    \centering
    \includegraphics[width=.95\textwidth]{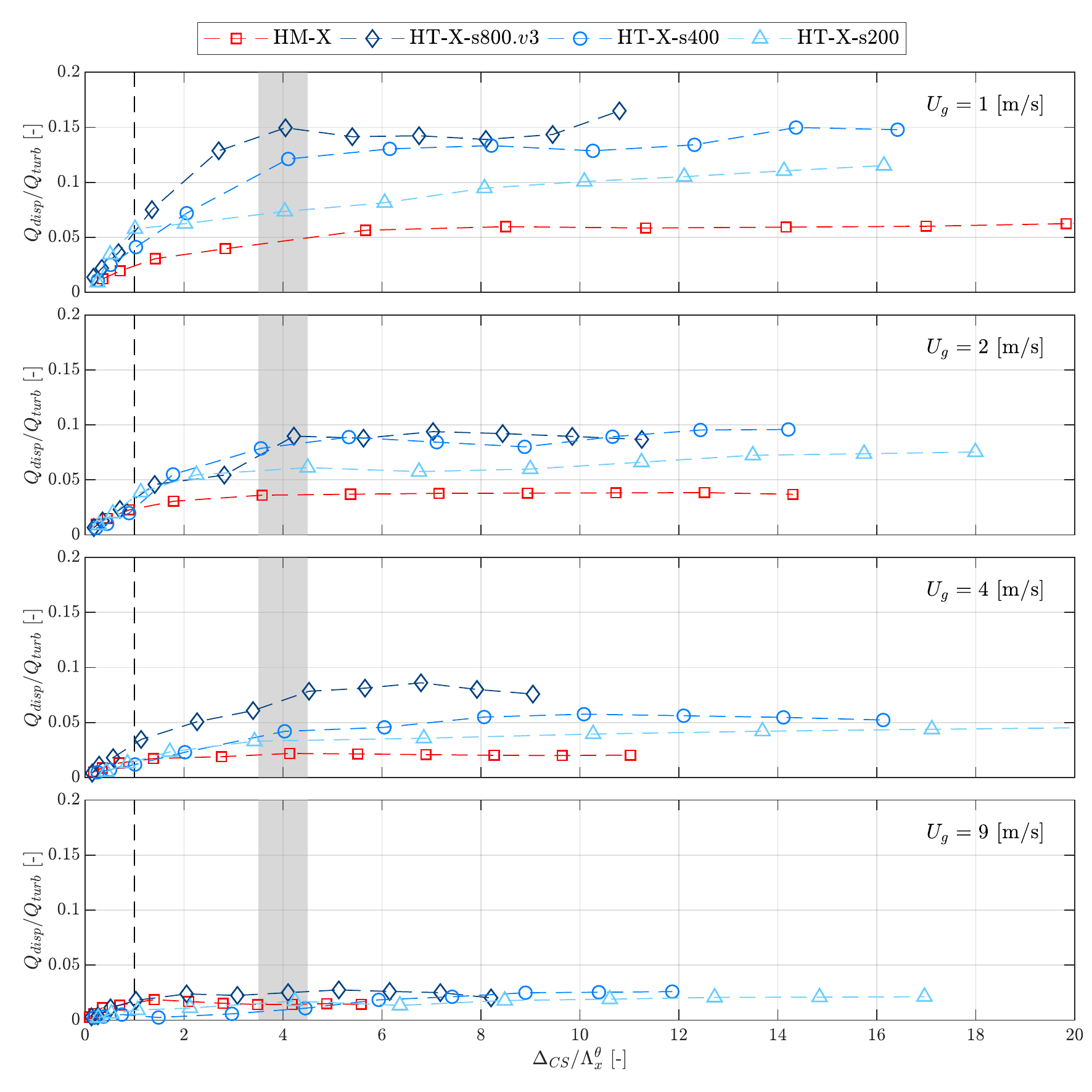}
    \caption{Ratio of the dispersive fluxes versus the turbulent fluxes as a function of averaging surface defined by the Control Surface area $\Delta_{CS}$ and the dispersive integral length scale for temperature $\Lambda_x^{\theta}$. From top to bottom, the geostrophic forcing increases. The dashed line illustrates where $\Delta_{CS}/\Lambda_{x}^{\theta} = 1$, and the grey shaded bar, the corresponding averaging surface area from whereon the contribution of the dispersive fluxes versus the turbulent fluxes reaches equilibrium.}
    \label{fig:CV}
\end{figure}

To address these questions, Figure \ref{fig:CV} provides a measure of the contribution of the dispersive fluxes versus the turbulent fluxes as a function of the horizontal averaging length scale, defined here by a control surface area ($\Delta_{CS}^2$) normalized by a `\textit{dispersive}' integral length scale. Results are obtained at the study height of 32 m. This height corresponds to the fourth grid point in the LES domain in the vertical direction, in which the potential numerical artifacts introduced by the LES wall-boundary conditions have been diffused, and also approximately corresponds to the height of the first grid point in NWP models. This is therefore where surface parametrizations would be applied in NWP models.     

The dispersive integral length scale ($\Lambda_x^{\theta}$, dispersive integral length scale for temperature in the x-direction) characterizes the footprint of the surface heterogeneities on the mean flow. This length scale is computed through the correlation of the spatial fluctuations of the time and space averaged temperature field ($\theta''$) in the x-direction, similar to what traditionally is done to compute the turbulent integral length scale \citep{Pope2000}. This dispersive length scale ($\Lambda_x^{\theta}$) is assumed equivalent in the following narrative to $l_h$ in Figure \ref{fig:LengthScales} for the surface characteristic heterogeneity length scale. Results indicate that when the ratio between the averaging control surface characteristic length scale $\Delta_{CS}$ and $\Lambda_x^{\theta}$ is about $\sim 4$ one obtains the full contribution of the dispersive fluxes, with only minor changes to the fluxes as $\Delta_{CS}$ is increased. 

\begin{figure}[t]
    \centering
    \includegraphics[width=.95\textwidth]{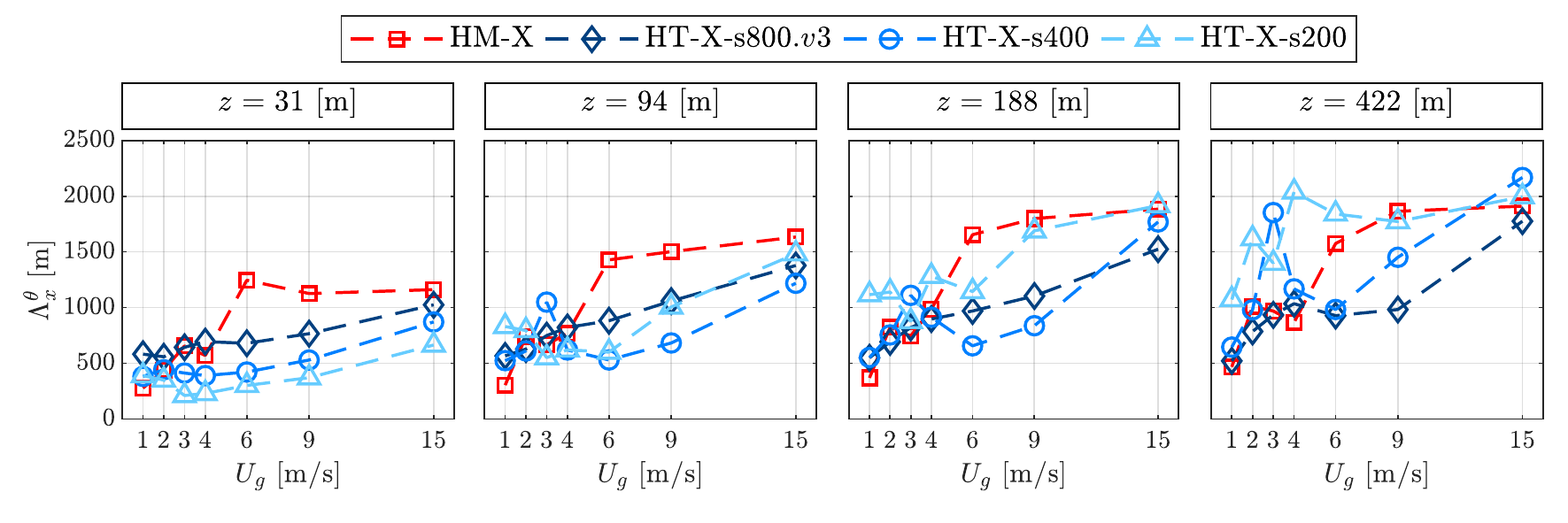}
    \caption{Values of the dispersive integral length scale for temperature in the streamwise direction $\Lambda_{x}^{\theta}$ as a function of increasing height (left to right) and increasing the geostrophic wind speed.}
    \label{fig:LengthScales2}
\end{figure}

Because, this dispersive integral length scale is of similar magnitude to the size of the surface thermal patchiness when the geostrophic forcing is weak, as indicated in Figure \ref{fig:LengthScales2}, even when considering four times the size of the largest patches, this measure remains smaller than the physical size of the numerical domain ($2\pi\,z_i$). Therefore, an immediate consequence of the results in Figure \ref{fig:CV} is that the results in Section \ref{sec:results} remain valid despite having been averaged over the full LES domain. Also, from Figure \ref{fig:CV}, it is worth noting that the contribution of the dispersive fluxes versus the turbulent fluxes tends to zero when the averaging control surface is small in comparison to the size of the thermal heterogeneity length scale, which is equivalent to the hypothesis of Figure \ref{fig:LengthScales} (\textit{i.e.}, when $l_e >> l_h$, turbulence feels the surface as homogeneous). Alternatively, when averaging over the correct length scales, the dispersive fluxes can contribute between $\sim 3$ to $15\%$ of the turbulent fluxes in moderate to low wind speeds. This result, although from a highly idealized scenario, already provides an initial preliminary response to the earlier hypothesis of whether dispersive fluxes could account for the 5-10\% of energy that is traditionally missing when computing surface energy budgets \citep{Foken2008,Stoy2013}. This result requires further confirmation either using experimental data or more realistic numerical simulations. 

\begin{figure}[t]
    \centering\includegraphics[width=.95\textwidth]{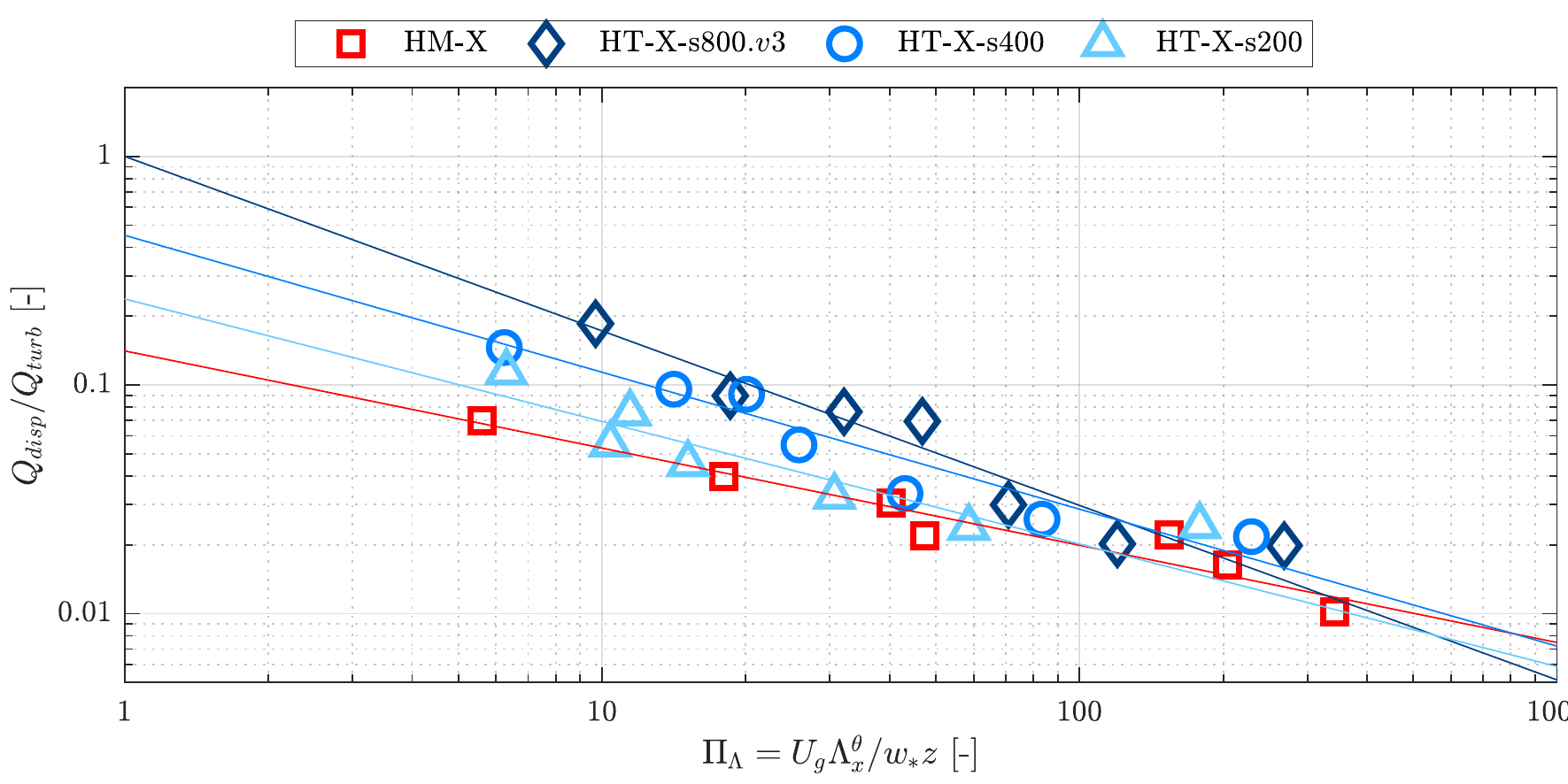}
    \caption{(a) Ratio of the dispersive fluxes versus the turbulent fluxes as a function of $\Pi_{\Lambda}$.}
    \label{fig:Re}
\end{figure}

To further investigate the relationship between the geostrophic forcing, the mean flow surface induced heterogeneities, and the corresponding near surface turbulence, we introduce a non-dimensional parameter representing the ratio of the large scale advection effects to convective turbulence,
\begin{equation}
    \Pi_{\Lambda} = \frac{U_g\,\Lambda_{x}^{\theta}}{w_*\,z}.
\end{equation}
Note that $U_g\,\Lambda_{x}^{\theta}$ and $w_*\,z$ have unites of $l^2/t$ which is similar to diffusivity or viscosity. As it can be observed in Figure \ref{fig:Re}, when the ratio of the dispersive fluxes versus the turbulent fluxes is plotted as a function of the new non-dimensional number, results illustrate a negative power law of the type

\begin{equation}\label{eq:disp}
\frac{Q_{disp}}{Q_{turb}} = \alpha\,\Pi_{\Lambda}^{-\beta}.
\end{equation}

In this relationship, the power coefficient ($\beta$) increases with the surface thermal patch size. Interestingly, when $\Pi_{\Lambda}$ is large ($\sim \mathcal{O}(1000))$, the corresponding contribution of the dispersive fluxes in the heterogeneous cases is similar to that of the homogeneous cases. It is in this regime that the near surface turbulence can not effectively transport the effects of the surface thermal heterogeneities up into the surface layer, and instead the measure of dispersive inertia is representative of turbulent coherent structures produced by shear and not by the thermal forcing from the surface. These results are consistent with those illustrated earlier in Figure \ref{fig:DispFluxes} as well as with the initial hypothesis formulated in Figure \ref{fig:LengthScales} which states that when the turbulent eddies ($l_e << l_h$) are too small the near-surface turbulence feels the actual surface as homogeneous and traditional ASL formulations remain valid (indicated here by the small contribution of the dispersive fluxes).  Also of interest is the limit of $\Pi_{\Lambda}$ tending to one (representing the case where $l_e \sim l_h$, as posed in Figure \ref{fig:LengthScales}) in which case the contribution of the dispersive fluxes is similar to that of the turbulent fluxes (\textit{i.e.} $\sim 25\%$ for HT-X-s200, $\sim 45\%$ for HT-X-s400, and $\sim 100\%$ for HT-X-s800). It is in this regime that the thermal heterogeneities play an important role in the energy exchange between the surface and the turbulent flow aloft, and results indicate that the dispersive fluxes could provide a means to capture this effect when heterogeneities are unresolved in NWP models. In the final limit case of $\Pi_{\Lambda} \rightarrow 0$, the problem reduces to the canonical Rayleigh-B\'{e}nard convection in which the dispersive inertia tends to zero. In this case the contribution of the dispersive fluxes is larger than the turbulent fluxes. Note that this special case should only be considered as a conceptual experiment, given that in practice situations in which pure Rayleigh-B\'{e}nard convection are encountered in real ABL flows are very limitted given the co-existence of other flow forcings. 

Finally, it is worth mentioning that this analysis has also been done at different heights (below the vertical height above which the contribution of the dispersive fluxes does not change by more than 10\%, see Figure \ref{fig:DispFluxes_Profiles}), with the main trends only  illustrating small changes. Also, note that the value of $\alpha$ in equation \ref{eq:disp} can take any value given that when $\Pi_{\Lambda} \rightarrow 0$, $Q_{disp}/Q_{turb} \rightarrow \infty$. For simplicity a value of $\alpha = 1$ is here selected.

\section{Discussion}\label{sec:discussion}

In the early 1980's, the concept of dispersive fluxes was introduced to account for the momentum fluxes arising in vegetated canopies induced by the interaction of the turbulent flow with the canopy structure and leading to persistent flow heterogeneities in time \citep{Raupach1982, Finnigan1985, Raupach1986}. In this work, dispersive fluxes are reinterpreted as a term arising from the spatial averaging operation that explicitly represents critical processes dependent on heterogeneity induced in the flow by surface thermal patchiness. This can be easily generalized to any process inducing persistent flow heterogeneities. From the analysis presented, two main results are extracted: the contribution of dispersive fluxes to the total energy exchange in the ABL can be important; dispersive fluxes can be associated with the topology of the surface heterogeneity, or with the persistent turbulent structure of the ABL flow depending on the intensity of the geostrophic forcing and the time averaging operation. Further, as originally expected, the ratio between the characteristic heterogeneity length scale $l_h$ and the energetic turbulent eddies $l_e$ is a controlling parameter of the contribution of energy flux captured by the dispersive fluxes. Therefore, while the resolution of the current and near future NWP models remains larger or close to the characteristic length scale of the most energetic eddies in the ABL ($\sim z_i$), and the unresolved heterogeneity is also of similar order, then dispersive fluxes can play an important role in describing the unresolved energy and momentum fluxes. Therefore, one could envision developing new parameterizations for the unresolved momentum and energy fluxes based on the structure of dispersive fluxes. Further, results also seem to indicate the existence of two regimes where the role of the dispersive fluxes is modulated by different effects:one in which dispersive fluxes are driven by surface heterogeneities; and, another regime where dispersive fluxes are driven by long-lived coherent structures of atmospheric turbulence in high-shear conditions. Therefore, differentiation based on these two regimes could facilitate developing new parameterizations. 

In the development of the different study cases, it is also important to realize that the simulations are forced through an imposed surface temperature, which eventually leads to different atmospheric stability values for the different study cases. While this could be an important limitation in a study that focused on a one-to-one intercomparison of cases, it does not affect the interpretation of the results in this specific work. This is because in here the aim is focused on illustrating that dispersive fluxes can be relevant in realistic ABL conditions, and that these are dependent on geostrophic forcing, heterogeneity length scale, and time averaging. The reason for forcing the flow with an imposed surface temperature, instead of surface flux as done for example in \cite{Salesky2016} is because the design of the simulations was inspired from recent experimental measurements using a thermal camera in Utah Salt Flats near the SLTEST site on the Dugway Proving Grounds, Utah \citep{Morrison2017}.

\section{Conclusions}\label{sec:conclusions}

In this work we present an LES study of the influence that surface thermal heterogeneities can have on the atmospheric boundary layer flow as a function of geostrophic forcing, and as a function of thermal patch size. For the first time, we propose the use of dispersive fluxes as a measure of the footprint that these surface thermal heterogeneities have on the flow. Results illustrate that, under weak geostrophic forcing, dispersive fluxes can account for up to 40\% of the total sensible heat flux at about $0.1z_i$, with a value of 5 to 10\% near the surface. These provide an indirect measure of the footprint that thermal heterogeneities have on the flow. Under stronger geostrophic forcing, heterogeneities are blended, changing the turbulent structure of the flow, and reducing the measured values for the dispersive fluxes $\sim 5\%$. In this later case dispersive fluxes provide a measure of the coherent structure of the turbulent flow induced by the ground surface shear stress. This character seems to be governed by a non-dimensional parameter representing the ratio of the large scale advection effects to convective turbulence.

\begin{acknowledgements}
This project has been developed with the support of the U.S. National Science Foundation grant number PDM-1649067. Marc Calaf also acknowledges the Mechanical Engineering Department at University of Utah for start-up funds, and the Center for High Computing Performance (CHPC) at University of Utah for computing hours. This work used the Extreme Science and Engineering Discovery Environment (XSEDE), which is supported by National Science Foundation grant number ACI-1548562. The authors declare no conflict of interest.
\end{acknowledgements}

\bibliographystyle{spbasic_updated}      

\end{document}